\documentclass{ar2e}

\usepackage{ulem}
\usepackage{ARAstroBib}
\usepackage{amssymb,amsbsy,psfig}
\usepackage{xspace}
\usepackage[usenames]{color}
\usepackage{graphicx}

\def\apj{ApJ}                                         
\def\apjs{ApJS}
\def\apjl{ApJL}
\def\aap{A{\&}A}

\def\mnras{MNRAS}
\def\aj{AJ}

\def\pasp{PASP}
\def\pasj{PASJ}
\def\nat{Nature}
\def\planss{Planetary Space Science}

\def\eg{{\it e.g.}\xspace}

\def\Sref#1{Section~\ref{#1}\xspace}
\def\Fref#1{Figure~\ref{#1}\xspace}

\def\CaseStudy#1{\noindent{\it\bf #1 \,\,\,\,}}
\def\url#1{\texttt{#1}}

\begin{document}


\jname{Annu.\ Rev.\ Astron.\ Astrophys.}
\jyear{2015}
\jvol{}
\ARinfo{}

\title{Ideas for Citizen Science in Astronomy}

\author{%
Philip J.\ Marshall
  \affiliation{Kavli Institute for Particle Astrophysics and Cosmology, P.O.~Box~20450, \newline
   MS~29, Stanford, CA 94309, USA.}
Chris J. Lintott
  \affiliation{Department of Physics, Denys Wilkinson Building, University of Oxford, \newline
   Keble Road, Oxford, OX1 3RH, UK.}
Leigh N. Fletcher
  \affiliation{Atmospheric, Oceanic and Planetary Physics, Clarendon Laboratory, University
   of Oxford, Parks Road, Oxford OX1 3PU}
}

\markboth{Marshall, Lintott \& Fletcher}{Citizen Astronomy}



\begin{abstract} 

We review the relatively new, internet-enabled, and rapidly-evolving field of
citizen science, focusing on research projects in stellar, extragalactic and 
solar system astronomy that have benefited from the participation of members of
the public, often in large numbers. We find these volunteers making
contributions to astronomy in a variety of ways: making and analyzing new
observations, visually classifying features in images and light curves,
exploring models constrained by astronomical datasets, and initiating new
scientific enquiries.  The most productive citizen astronomy projects involve
close collaboration between the professionals and amateurs involved, and occupy
scientific niches not easily filled by great observatories or machine learning
methods: citizen astronomers are most strongly motivated by being of service to
science. In the coming years we expect participation and productivity in citizen
astronomy to increase, as survey datasets get larger and citizen science
platforms become more efficient. Opportunities include engaging the public in
ever more advanced analyses, and facilitating citizen-led enquiry by designing
professional user interfaces and analysis tools with citizens in mind.

\medskip The making of this review is still in progress. The most up to date PDF
file can be downloaded from github at 
\url{http://tinyurl.com/CitizenAstronomyReview} We invite feedback on the review
via its repository's {\it issues} at 
\url{http://tinyurl.com/CitizenAstronomyReviewFeedback} and aim to submit to
ARAA on September 26.

\end{abstract}

\maketitle


\section{INTRODUCTION}
\label{sec:intro}

The term ``citizen science'' refers to the activities of people who are not paid
to carry out scientific research (``citizens''), but who make intellectual
contributions to scientific research nonetheless.\footnote{In this review we
differentiate between the data analysis that citizens carry out themselves, and
distributed ``grid'' computing farmed out to processors owned by citizens, and
omit the latter since it does not fit our definition of citizen science as
involving intellectual  contributions from citizens.} They come from all walks
of life, and their contributions are diverse, both in type and research area.
This review is about the astronomy projects they have participated in to date,
the tasks they have performed, and how astronomy has benefited -- and could
benefit further -- from their efforts.

Citizen involvement in science pre-dates the profession itself. The earliest
example of collaboration between professional and amateur astronomers seems to
have been Edmund Halley's call for observations of the 1715 total eclipse of the
Sun which crossed central England \citep{Halley}.\footnote{The aim was to refine
estimates of the size of the shadow cast by the Moon, and citizen observations
were much needed. Although Halley was successful in observing, his colleagues at
Oxford were clouded out, and those in Cambridge were ``oppressed by too much
Company, so that, though the heavens were very favourable, [they] missed both
the time of the beginning of the Eclipse and that of total darkness.''}  Since
then there has been a long and honourable tradition of amateur observers
making important discoveries and significant sustained contributions. However,
the advent of the world wide web has changed the face of professional and
amateur collaboration, providing new opportunities and accelerating the sharing
of information. People are now connected to each other on a scale that has never
happened before. Professional scientists can interact with citizens via a range
of web-based media, including purpose-built citizen science websites which
increase the potential for shared data analysis and exploration, as well as for
data collection. Meanwhile, communities of citizens have sprung into existence
as like-minded people have been able to find and talk to each other in a way
that is almost independent of their geographical location. The result has been
an exponential increase in citizen involvement in science. The field is evolving
very quickly, with more and more professional scientists becoming aware of the
possibilities offered by collaborating with, for example, specialists operating
outside the usual parameters of professional astronomical observation, or tens
of thousands of people eager to perform microtasks in their spare time.  

Our aim in this work is to review the astronomical (and occasionally wider)
literature for productive citizen science projects, and distill the
characteristics that made these case studies successful.  As our title states,
this is a review of ideas for astronomy: we will look forward as well as back,
and try to answer the following questions. What are the particular niches that
citizen science fills, in our field? What is the potential of citizen science in
astronomy, and how can it be realized?  Citizen science has a significant impact
on its participants,  whether they be sitting in a university office or in front
of a home computer or mobile phone screen. This review is about the impact that
citizen astronomy has had, and can have, on the progress of  research in
astronomy.

This review is organised as follows. Astronomy research typically starts with
observations: so do we, in \Sref{sec:obs}. We then proceed to consider visual
classification, data modeling and finally citizen-led enquiry in 
Sections~\ref{sec:class}--\ref{sec:explore}. With this overview in place, we
take a look in \Sref{sec:crowd} at the population of citizens who take part in
astronomical research. We then turn to the future, and speculate on how citizens
might contribute to astronomy there (\Sref{sec:future}), and finish with some
concluding remarks in \Sref{sec:conclusions}.


\section{AMATEUR OBSERVING}
\label{sec:obs}

There is currently an active community of well-equipped amateur observers making
astronomical observations of great utility. There are also many other citizens
observing the night sky with less sophisticated equipment -- and as we shall
see, there are even some examples of citizens making astronomical observations
almost inadvertently. What astronomical data are the citizenry taking, and what
is it being used for?


\subsection{Active Observing}
\label{sec:obs:active}

In this section, we review some of the citizen contributions to active
observations of the night sky.  ``Passive'' contributions  will be described in
\Sref{sec:obs:passive} below. The steady improvements and increasing
affordability of digital technology, in addition to the ease of data sharing and
communications, have considerably expanded the realm of amateur astronomy in the
past two decades.  Meanwhile, professional observatories are always
over-subscribed, with resources necessarily being divided between particular
areas of sky, or samples of objects, or on a few astronomical questions: tuning
the parameters of professional observations to optimize all possible scientific
enquiries would seem an impossible task. What types of  niches does this leave
for amateur observers to fill? What are the strengths that amateur observers can
play to?

The first key advantage that amateurs have is time availability. Determinations
of meteor frequencies (for example) require observations on short timescales
(minutes), whereas the slow evolution of giant planets (for example) occurs on
longer timescales (years and decades).  Amateur observations can be frequent and
repetitive, but also long standing. 

The second, related, advantage is that of flexibility: whenever a new phenomenon
is discovered (e.g., a new comet, or anything changing the appearance of the
familiar planetary discs), observers will be keen to catch a glimpse
irrespective of the scientific value of their observations.  This reaction can
be near instantaneous, compared to the need to allocate telescope resources
among the professional community, and, when made by a networked community,
provides naturally well-sampled coverage across the globe.

The third benefit is contextual.  Professional observations are often taken in a
very different wavelength range, focus on a narrower spatial region, or employ
spectroscopic techniques that don't yield images. In some situations,
near-simultaneous wide field optical imaging by citizen scientists can provide
useful additional constraints on the process of interest.

The example case studies below serve to illustrate this synergy between amateur
and professional observations, and also to highlight instances of
professional-amateur (``Pro-Am'') collaboration. While the solar system provides
some of the most amenable targets for amateur observation, ``deep sky''
observations by the non-professional community provide important further insight
into the capabilities and potential of citizen astronomers.


\CaseStudy{Discovery and characterisation of asteroids and comets.}
Although survey telescopes provide the vast majority of modern solar system
discoveries, citizen astronomers occupy some useful observational niches.  Small
solar system objects moving against the fixed-star background can be detected in
a set of CCD frames either by eye or by automated software. Targets include
near-earth asteroids (NEAs, with orbits intersecting those of the terrestrial
planets), main belt asteroids between Mars and Jupiter, and comets making their
journey towards the Sun from the outer solar system. The extreme familiarity of
citizen astronomers with a particular region of sky, planet or nebula, allows
them to immediately identify peculiarities or new features.  A protocol for
citizen discovery has been established: the position of any new object is
compared to existing catalogues, and if no existing details are found then the
new discovery and its ephemerides can be reported to the IAU Minor Planet
Center.\footnote{\url{http://www.minorplanetcenter.net}} If observations are
repeated for at least two nights by one or several observers, then a new
denomination is provisionally assigned to the discovery. An electronic circular
then reports the discovery to the wider world. For example, the NEA 2012~DA14
was initially reported to the Minor Planet Center by a team of amateur observers
affiliated with the La Sagra Sky Survey at the Astronomical Observatory (Spain),
and subsequently  characterised by professional astronomers during its closest
approach in February 2013 \citep[e.g., ][]{13deleon}.

As with asteroids, the majority of new comet discoveries are made by automated
surveys, but a small and stable number of discoveries come from amateurs with
small telescopes, typically in regions poorly covered by survey telescopes
(e.g., regions close to the Sun).   C/2011~W3~(Lovejoy), a Kreutz sungrazer
comet, is one such example, discovered by T. Lovejoy and circulated via the
Central Bureau for Astronomical Telegrams (CBAT) \citep[e.g.,][]{12sekanina}. 
The Oort cloud comet C/2012~S1~(ISON) was spotted by V. Nevski and A. Novichonok
in images from the International Scientific Optical Network, which spurred a
major international effort to observe its perihelion passage as it disintegrated
\citep{14sekanina}.  Amateurs are also contributing to the search for a
sub-category of objects with a detectable cometary coma within the asteroid
belt.  Recent discoveries of these main belt comets, which appear to be
asteroids that are actively venting their volatiles at perihelion, are beginning
to blur the distinction between asteroids and comets.  The T3 project, a
collaboration between the University of Rome and several amateur observers,
began in 2005 with the detection of a coma around asteroid 2005~SB216
\citep{06buzzi}, and has gone on to detect at least eight main belt comets
\citep{14mousis_proam}.  These early citizen science discoveries, followed up by
professional astronomers, have generated new insights into the properties and
variety of comets, and the dynamic and evolving nature of our solar system.  The
discovery of Comet Shoemaker-Levy 9 (co-discovered by amateur observer D. Levy)
before its collision with Jupiter \citep{04harrington} is a classic example.
In general, it is the global distribution of citizen observers and the
long-baselines of their observations that enable new discoveries of minor bodies
in our solar system.

Beyond first detections, citizen observers can aid in the detailed study of the
physical and orbital characteristics of these newly discovered solar system
bodies. These amateur-led contributions are typically published via the
\textit{Minor Planet
Bulletin}.\footnote{\url{http://www.minorplanet.info/mpbdownloads.html}} 
Photometric monitoring of an asteroid as it rotates provides information on its
physical parameters such as its shape, rotation rate and orientation; 
monitoring of a comet's coma, dust and plasma tails can reveal dynamic
structures, determine the locations of active venting regions and reveal
outbursts and other events associated with the outgassing \citep[see][for a
comprehensive review]{14mousis_proam}.  

\begin{figure}[!ht]
\centering\includegraphics[width=\linewidth]{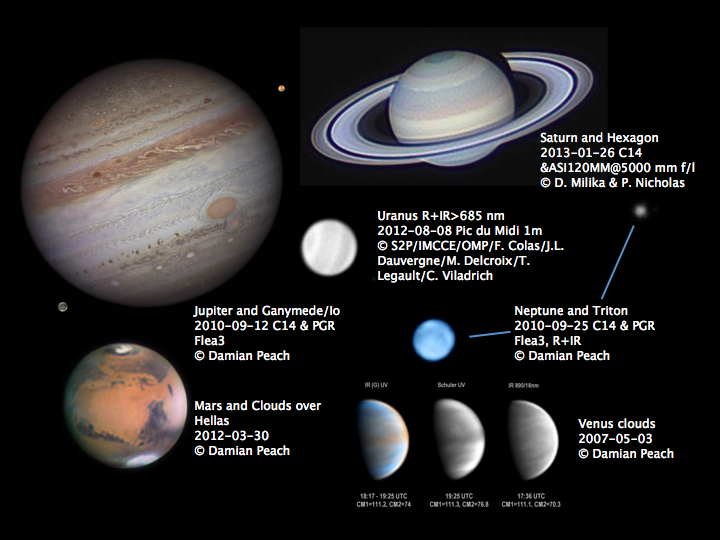}
\caption{Examples of high fidelity images obtained by amateur planet
observers.}
\label{fig:planets}
\end{figure}

\CaseStudy{Planetary monitoring over long timescales}
Planetary atmospheres make tantalising targets for citizen observers, being
large, bright, colourful and highly variable from night to night (e.g., 
\Fref{fig:planets}.  The long-term monitoring provided by the network of amateur
astronomers provides valuable insights into the meteorology and climate of these
worlds, tracking the motions of clouds, waves and storms as they are transported
by atmospheric winds to probe the physical and chemical processes shaping their
climates.  For example, the global distribution of giant planet observers
permits global monitoring of Jupiter and Saturn as they rotate over 10 hours.
Citizens upload raw filtered images and colour composites, organised by date and
time, to online servers, such as the Planetary Virtual Observatory and
Laboratory (PVOL\footnote{\url{http://www.pvol.ehu.es/pvol}}) maintained for the
International Outer Planets Watch \citep[IOPW][]{10hueso}.  Those images can be
used by amateurs and professionals alike to study quantitatively the visible
activity, including  measuring wind speeds from erupting plumes
\citep{08sanchez}, investigating the strength and changes to the large vortices
\citep[e.g., the 2006 reddening of Jupiter's Oval BA,][]{06simon-miller}, and
determining the life cycle of the belt/zone structure \citep{96sanchez,
11fletcher_fade}.  For Saturn, a close collaboration between citizen scientists
and Cassini spacecraft scientists (known as Saturn Storm Watch) has allowed
correlation of lightning-related radio emissions detected by the spacecraft with
visible cloud structures on the disc \citep[e.g.,][]{11fischer}, which would not
have been possible with the targeted regional views provided by Cassini's
cameras alone. Furthermore, it was the amateur community that first spotted the
eruption of Saturn's enormous 2010-2011 storm system, which was monitored over
several months \citep{12sanchez}.

Video monitoring has been used by citizen observers to enable high resolution
``lucky'' imaging of Jupiter. The best images, at moments of clear seeing, from
the high-resolution video frames are selected, extracted and stacked together,
using custom software to correct for the distortions associated with the
telescope optics and residual atmospheric seeing.  Software written by citizen
scientists for free distribution to active observers, such as
Registax\footnote{www.astronomie.be/registax} and
Autostakkert\footnote{www.autostakkert.com},  allows them to process their own
video files, thus avoiding the need for transfer of large datasets to some
centralised server \citep[see][for a thorough review]{14mousis_proam}. 
Descriptive records of morphological changes are maintained and continuously
updated by organisations of citizen scientists such as the British Astronomical
Association (BAA) and the Association of Lunar and Planetary Observers (ALPO).
The BAA's Jupiter section\footnote{\url{http://www.britastro.org/jupiter}} is a
team of amateurs with substantial expertise in Jupiter's appearance
\citep{95rogers};  their regular bulletins describe the changing appearance of
the banded structure and the emergence of new turbulent structures and weather
phenomena, and keep a record of the long-term atmospheric changes.  

Active citizen observing also provides long-term monitoring in the inner solar
system.  Venus' photochemical smog shields the planet's surface from view, but
discrete cloud features can be used to study the super-rotation of the Venusian
atmosphere, and the occurrence of a mysterious ultraviolet absorber at high
altitudes.  For example, the Venus Ground-Based Image Active Archive was created
by ESA to provide contextual observations supporting the Venus Express mission
\citep{08barentsen}.  The Martian atmosphere, with its ephemeral clouds,
seasonal CO$_2$ polar ice cycles, and dust storms, continues to prove popular
among citizen observers, although these are a minor supplement to the wealth of
high-resolution information being returned by orbital and surface missions to
the red planet.  As with other planetary targets, amateur observations provide
the long temporal records of the evolution of atmospheric features.  Groups such
as the International Society of Mars Observers
(ISMO\footnote{\url{http://www.mars.dti.ne.jp/~cmo/ISMO.html}}), the British
Astronomical Association (BAA) and the International Mars Watch program
quantitatively and qualitatively assess these amateur images.  Finally, although
citizen observations of Uranus and Neptune are in their infancy and require
telescopes with diameters exceeding 25 cm, there have been confirmed reports of
atmospheric banding and discrete cloud features when near-infrared filters  (to
maximise the contrast between the white clouds and the dark background) and long
exposure times of tens of minutes are used.  Citizen monitoring of all of these
worlds (summarised in \Fref{fig:planets}) provides the long-baselines,
flexibility and high frequency of imaging needed to understand the forces
shaping their evolving climates.

\begin{figure}[!ht]
\centering\includegraphics[width=\linewidth]{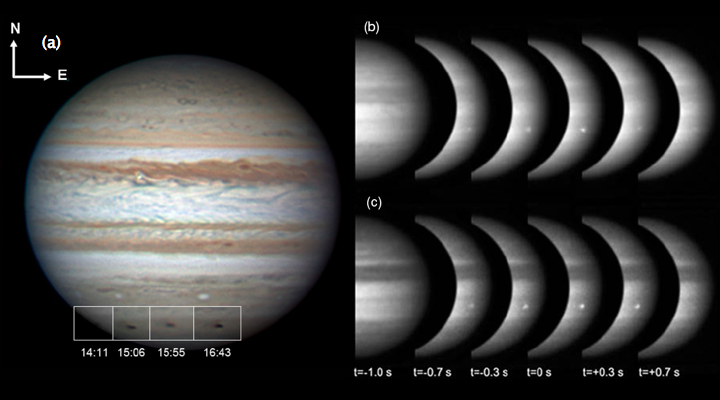}
\caption{Citizen science contributions to monitoring of impacts in the Jupiter
system. (a) Dark impact scar in Jupiter's atmosphere imaged by Anthony
Wesley on July 19th 2009 \citep{10sanchez}. (b) The
evolution of a smaller bolide impact on June 3rd 2010 at red
wavelengths, also imaged by Wesley. (c) The evolution at blue
wavelengths by Christopher Go, figure from \citet{10hueso}.}
\label{fig:jupiter-impacts}
\end{figure}

\CaseStudy{Solar System Impacts.}
The increasing adoption of video monitoring of planetary targets means that
unexpected, short-lived events on the surfaces of those bodies 
are now more likely to be observed by citizen
astronomers than by professional observatories.  For example, an impact scar
near Jupiter's south polar region was first discovered in imaging by Australian
amateur Anthony Wesley on July 19th, 2009. This led to an international campaign
of professional observations to understand the asteroidal collision that had
created the scar \citep[e.g.,][]{10hammel,10depater,11orton}.  Although the 2009
impact was out of view from the Earth, at least three flashes have been
confirmed between 2010 and 2012, and the light curves used to determine the
sizes and frequency of objects colliding with Jupiter \citep[e.g.,][]{10hueso}
(\Fref{fig:jupiter-impacts}).  Citizen scientists have developed free software
to allow observers to search for impact flashes in an automated way (e.g.,
Jupiter impact detections\footnote{\url{http://www.pvol.ehu.es/software}} and
LunarScan from the ALPO Lunar Meteoritic Impact Search for transient impact
flashes recorded on the moon
\footnote{\url{http://alpo-astronomy.org/lunarupload/lunimpacts.htm}}).  


\CaseStudy{Transiting and Microlensing Exoplanets.}  
Amateur observers have contributed to several exoplanet investigations,
responding to detections made by professional surveys and making important
contributions to the light curves of the targets. In the case of exoplanet
transits, the challenge is to measure the 1\% diminution in starlight as a giant
planet transits in front of its parent star. \citet{14mousis_proam} point out three
methods whereby amateurs can contribute to the characterisation of exoplanetary
systems: first, by frequent observations of known transits to refine ephemeris;
second, by searching for transit time variations that can reveal additional
planets in a system; and third, by searching for previously unidentified
transits in known planetary systems \citep[e.g., the discovery of the transit of
HD 80606b from a 30 cm telescope near London,][]{09fossey}. In a planetary
microlensing event, significant brightening of the background star is required
to make a planet orbiting the microlens visible at all; if additional caustic
crossings are caused, the resulting exoplanetary microlensing feature is of just
several days duration, calling for high frequency, on demand monitoring -- a
situation well matched to the capability of a global network of small telescope
observers \citep[see e.g.\ ][]{Christie2006}. High magnification events
detected by the  OGLE\footnote{\url{http://ogle.astrouw.edu.pl/}} and
MOA\footnote{\url{http://www.phys.canterbury.ac.nz/moa}} surveys have been
broadcast by the
microFUN\footnote{\url{http://www.astronomy.ohio-state.edu/~microfun}} and
PLANET\footnote{\url{http://planet.iap.fr}} networks (now merged) to
globally-distributed professional and amateur observers to follow up. These
collaborations have been very successful, helping enable characterisation of
over a dozen exoplanet systems \citep[see e.g.\ ][and references
therein]{Udalski++2005,Gould++2014}.

%
%

\CaseStudy{Variable Star Monitoring: the AAVSO.}
The American Association of Variable Star Observers (AAVSO) supports and
coordinates the efforts of over 900 amateur observers who are interested in
bright, nearby variable stars; in 2013, the community made over a million
observations, either visually or with CCD or DSLR cameras, and logged them
into a shared database.\footnote{The AAVSO annual reports can be found at
\url{http://www.aavso.org/annual-report}} The AAVSO provides a number of
services to assist the volunteers, including training material, an online data
entry tool that carries out basic error checking, several pieces of
software to assist the observers in checking their own observations, and data
reviews by AAVSO staff. Despite its name, AAVSO observers are located all over
the world, with two thirds of the membership working outside the US. Some of
the community's larger telescopes can be operated robotically, and have been
linked together into a network, AAVSOnet. This network is engaged in an
ongoing all sky survey, APASS,\footnote{\url{http://www.aavso.org/apass}}
which is carrying out a survey of over 50 million stars between 10th and 17th
magnitude, in 5 optical filters ($BVg'r'i'$). The data processing and
calibration is being done as a Pro-Am collaboration, and the data is being
released at approximately annual intervals.

The distributed nature of the AAVSO community means that it can produce
continuous light curves for stars at a wide range of declinations. The AAVSO
data has been used extensively by professional astronomers needing the most 
up to date optical measurements of stellar variability in, for example, the 
SS Cyg system \citep{Miller-Jones++2013},  optical light curves taken
simultaneously with monitoring being carried out by space telescopes and/or at
different wavelengths \citep[see \eg][for a successful joint AAVSO--HST
program]{Szkody++2013},  or who are performing long baseline data mining
analyses of variable star populations. 

The AAVSO, in partnership with several professional astronomers and education
specialists, successfully coordinated the NSF-funded ``citizen sky'' project
to monitor the 2009--2011 eclipse in the espilon Aurigae binary star system.
The results from this campaign  \citep{Stencel2012}\footnote{The results from
the Citizen Sky project are presented in a special issue of the JAAVSO at 
\url{http://www.aavso.org/jaavso-v40n2}} were used by \citet{Kloppenborg++2010}
to help interpret their interferometric imaging of the obscuring disk in the
system.



\subsection{Passive Observing}
\label{sec:obs:passive}

While amateur astronomers have aquired a great deal of very useful data, the
general population is better equipped than ever to image the sky and make that
data available for scientific analysis. This has been demonstrated by two
recent professionally-led studies, that made use of a largely passive
observing community connected via online social networks not usually
associated with astronomy. 


\CaseStudy{The Orbit of Comet Holmes from the Photographs Uploaded to Flickr.}
\citet{Lang++2012} used more than 2000 images scraped from the photo sharing
website Flickr as inputs to a reconstruction of the orbit of Comet Holmes. This
comet was bright enough to be visible with the naked eye during its 2007
apparition, and a large number of photographs were taken of it and uploaded to
the Flickr site. \citeauthor{Lang++2012} were able to astrometrically calibrate
many of the images using their automatic image registration software,
\texttt{astrometry.net} \citep{Lang++2010}, which detects the stars in each image and matches them
to those in professionally-assembled catalogs. This had been enabled as a Flickr
``bot,'' crawling over all images submitted to the \texttt{astrometry.net}
group\footnote{\url{https://www.flickr.com/groups/astrometry}} and sending the
photos' owners messages showing them where on the sky their images were taken. 
The time of observation could in many cases be derived from metadata included in
the image headers. The calibrated images trace out the trajectory of the comet,
producing a result which is close to that obtained from the JPL Horizons system
\citep{Giorgini}. Estimates of orbital parameters from Flickr images alone are
accurate, when compared to the Horizons values, to within a few standard
deviations. As the authors point out, while in this case the photographers did
not realize they were participating in a scientific study, the potential of
combining powerful calibration software with large amounts of citizen-supplied
imaging data is made clear. This method of ``unconscious'' citizen science may
prove to have significant value in fields beyond astronomy too, if models of the
statistical sampling can be developed, with ecological surveys of images
submitted to sites like Flickr likely in the next few years. 


\CaseStudy{Informal Earth-Meteor Impact Detection.} 
As well as observing other planets for impacts, citizen scientists have also
played a crucial role in the recording of asteroid impacts on Earth, although
not always realizing the usefulness of their observations. Video footage of
the fireball and shockwave of the February 2013 Chelyabinsk meteor
\citep{13popova} were essential to scientifically characterise the impactor and
its likely origins, despite the fact that these records were largely captured
accidentally by autonomous security cameras.  Trajectories reconstructed from
these records even permitted the recovery of meteorites from a debris field on
the ground.  

Such objects are fragments of comets and asteroids, the debris left over from
the epoch of planetary formation: their numbers, sizes and composition provide a
window into the earliest evolutionary stages of our solar system.  The
statistics of these impacts reveal the risk of threats in our local
neighbourhood, and these statistics are currently actively provided via a global
network of citizen scientists, sharing and publicising their observations of
meteors via the International Meteor Organisation
(IMO\footnote{\url{http://www.imo.net}}). However, visual observations of
meteors can also be tracked with no such active participation. By searching the
archive of short text messages submitted to the web service Twitter, Barentsen
et al (priv.\ comm.) were able to  detect several new meteor showers. Naked-eye
observers had spotted shooting stars and tweeted about them to their followers,
giving rise to a detectable signal in the steam of tweets that night. 



\section{VISUAL CLASSIFICATION}
\label{sec:class}

Observing the night sky with a telescope is perhaps the most familiar of the
activities of amateur astronomers, but as the previous section showed, citizens
are also actively involved in the processing and interpretation of the data they
have taken.  In this and the next section we look at projects where much larger
archival astronomical datasets have been made available to crowds of citizens,
who are asked to inspect images and light curves, and help describe and
characterize the features in them. Despite significant advances in machine
learning and computer vision, the visual inspection of data remains an important
part of astronomy, as it continues to take advantage of the amazing human
capacity for visual pattern recognition. While many in the 1990s predicted that
the increasing size of astronomical datasets would make such time-intensive
inspection impossible, the extensive reach of the world wide web has enabled the
involvement of hundreds of thousands of citizen scientists in this form of
``crowd-sourced'' data analysis. 


\subsection{Crowd-sourced Classification in Astronomy}
\label{sec:class:astro}

\CaseStudy{Stardust@home}
While significant preliminary work had been carried out by NASA's
``clickworkers'' (see below), the project that first illustrated the potential
of crowd-sourcing for astronomical purposes was 
Stardust@home\footnote{\url{http://stardustathome.ssl.berkeley.edu/}}. The team
asked volunteers to scan through images of samples returned from Comet Wild-2 by
the \emph{Stardust} mission, attracted a large audience to the apparently
unprepossessing task of looking for dust grains in an effort to identify samples
of material from outside our Solar System. The site was built on BOSSA, an early
attempt to build a generalized platform for such crowd-sourcing projects, and
featured a stringent test which volunteers had to pass before their
classifications would be counted. Despite this hurdle, more than 20,000 people
took part, and a variety of dust grains were removed from the aerogel for
further study, contributing two of the seven candidate interstellar grains presented in a recent Science paper\citep{Westphal}. Perhaps the most significant long-term impact of
Stardust@home, though, was the demonstration that large amounts of volunteer
effort were available even for such seemingly uninspiring tasks such as hunting
dust grains in images unlikely to be described as intrinsically beautiful, and
that, with a suitable website design and stringent testing, scientifically
valuable results could be obtained. 



\CaseStudy{Galaxy morphology with Galaxy Zoo}  

The Stardust@home experience directly inspired the development of Galaxy Zoo,
perhaps the most prominent scientific crowd-sourcing project to date. Galaxy Zoo
was built on the continued importance of morphological classification of
galaxies, first introduced in a systematic fashion by Hubble, and later
developed by, among others, de Vaucoleurs.  While the morphology of a galaxy is
closely related to its other properties, such as colour, star formation history,
dynamics, concentration and so on, it is not entirely defined by then: there is
more information in resolved images of galaxies than is captured in these
observables.  One approach was to develop simple proxies (e.g CAS \citep{Conselice}), but these
are at best approximations for true morphology. 

In an effort to prepare for large surveys, such as the Sloan
Digital Sky Survey (SDSS),
\citeauthor{Lahav1995}~(\citeyear{Lahav1995},\citeyear{Lahav1996}), and later,
\citet{Ball} developed neural networks trained on small samples of expert
classified images,\footnote{The Lahav papers are perhaps as interesting for
their psychology as for their astrophysics, as the classifications reveal the
relations between the senior classifiers employed to be experts.} in order to
automate the process of classification arguing that the size of the then
up-coming surveys left no place for visual classification.

The performance of these automatic classifiers depended on the input parameters,
including colour, magnitude and size. These variables correlate well with
morphology, but are not themselves morphological, and when included they
dominate the classification. In particular, for galaxies which do not fit the
general trends, such as spirals with dominant bulges, or star-forming
ellipticals, automated classifiers, whether using these simple measures or
more complex proxies for morphology such as texture, fail to match the
performance of expert classifiers\citet{Lin++2008}. As a result, \citet{Scha2007},
\citet{Nair}, and others have spent substantial amounts of time visually
classifying tens of thousands of galaxies. 

\begin{figure}[!ht]
\centering\includegraphics[width=\linewidth]{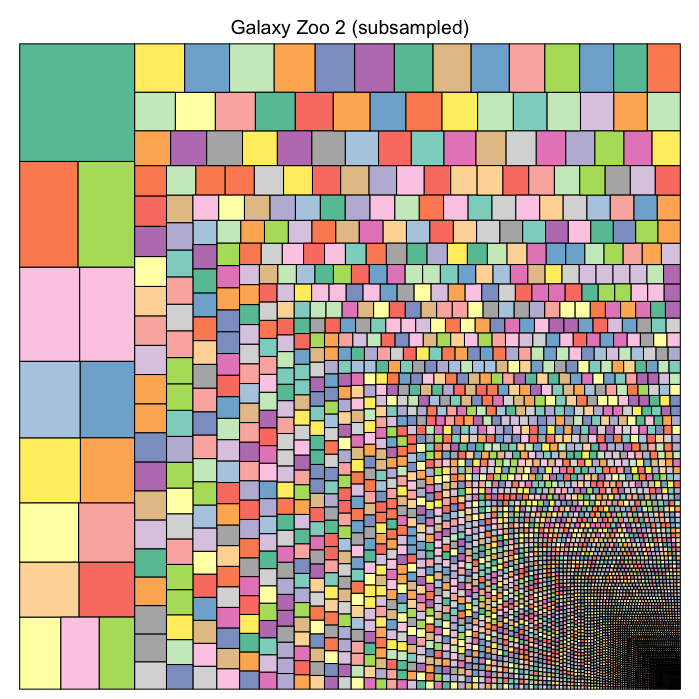}
\caption{Distribution of effort amongst 5000 randomly selected volunteers from
Galaxy Zoo 2. The area of each square represents the classifications of a single
user; colours are randomly assigned. The diagram illustrates the importance of
designing for both committed and new volunteers as both contribute significantly.}
\label{fig:gz2}
\end{figure}

Inspired by Stardust@home, a small group led by one of the authors (Lintott) created
Galaxy Zoo in 2007 to provide basic classifications of SDSS
galaxies\footnote{The original Galaxy Zoo is preserved at
\url{http://zoo1.galaxyzoo.org} with the current incarnation at
\url{http://www.galaxyzoo.org}.} Classifiers were presented with a coloured
image centered on and scaled to one of more than 800,000 galaxies, and could
select from one of six options to characterise that object's morphology: 
clockwise, anti clockwise and edge-on spirals,
ellipticals, mergers and ``star/don't know.'' Aside from  an easily-passed
initial test, little knowledge was required or indeed presented to classifiers,
enabling them to proceed quickly to doing something real soon after arriving at the
site; this approach, in contrast to Stardust@home, was
successful in encouraging large numbers of visitors to participate. 
This tactic -- in which both passing and sustained engagement provide
substantial contribution -- is illustrated in \Fref{fig:gz2} which shows results from
Galaxy Zoo 2. This later version of the project asked for more detailed
classifications via a decision tree containing questions such as `How prominent
is the bulge?', and later iterations of the project have applied a similar
approach to galaxies drawn from \emph{Hubble Space Telescope} surveys including
\textsc{GEMS, GOODS, COSMOS} and \textsc{CANDELS}.

To date, several hundred thousand people 
have participated in the Galaxy Zoo project, but such
figures would be meaningless if the classifications provided were not suitable
for science. With sufficient effort to ensure each galaxy is classified multiple
times (as many as 80 for many Galaxy Zoo images), these independent
classifications need to be combined into a consensus. As discussed in later
sections, this can become complex, but for Galaxy Zoo a simple weighting which
rewards consistency, first described in \citet{Land++2008}, was deemed sufficient.
Importantly, combining classifications provides not only the assignment of a
label but, in the vote fraction in a particular category, an indication of the
reliability of the classification. This allows more subtle biases, such as the
propensity for small, faint or distant galaxies to appear as elliptical
regardless of their true morphology, to be measured and accounted for
\citep[see][]{Bamford++2009}. The net result is that the Galaxy Zoo classifications
are an excellent match for results from expert classification, and have produced
science ranging from studies of red spirals \citep{Masters++2010} to investigations
of spiral spin \citep{Slosar++2009}.
A full review of Galaxy Zoo science is beyond the scope of this review; a recent
summary is given in the Galaxy Zoo 2 data release paper by \citet{Willett++2013}.

It is worth noting that some of the project's most important results
have been the result not of interaction with the main classification interface,
but represent rather serendipitous discoveries made by participants. 
We return to these in \Sref{sec:explore} below.



\CaseStudy{Surfaces of solar system bodies: Moon Zoo, Moonwatch.}
If studying galaxies remains, at least in part, a visual pursuit, then the same
is certainly true of planetary science. NASA's Clickworkers\footnote{http://www.nasaclickworkers.com/}, which asked
volunteers to identify craters on the Martian surface, lays claim to be the
oldest astronomical crowd-sourcing project. The consensus results matched those
available from experts at the time, but failed to go beyond this promising start
to produce results of real scientific value. More recently, interfaces inviting
classifiers to look at the Moon, Mercury, Mars and Vesta have been launched and
attracted significant numbers of classifications; however, although preliminary
results have been promising\citep{Kanefsky} these projects have yet to produce datasets that
have been used by the planetary science community in the same way that Galaxy
Zoo has by the astronomical community. The recent release of the first paper from the Cosmoquest
Moon Mappers project \citep{Robbins} may indicate that this will change.

\CaseStudy{Tracking Features in Giant Planet Atmospheres: WinJUPOS}
Not all astronomical crowd-sourced visual classification is led by 
professional scientists.
JUPOS\footnote{\url{http://jupos.privat.t-online.de}} is an amateur astronomy
project involving a global network of citizen observers to monitor the
appearance of planetary atmospheres.  Recent software developments have provided
a much more quantitative perspective on these citizen observations. The WinJUPOS
software\footnote{\url{http://jupos.privat.t-online.de}} was developed by a team
of citizen scientists led by G. Hahn; it allows multiple images of a giant
planet to be stacked with a correction for the rapid rotation of Jupiter or
Saturn  (once every 10 hours), then re-projected onto a latitude-longitude
coordinate system, so that the precise positional details of atmospheric
features can be determined via ``point-and-click,'' relying on the citizen's
ability to identify features on the planetary disc visually.  

By doing this over many nights surrounding Jupiter's opposition, the community
builds up enormous drift charts, comprising tens of thousands of positional
measurements, for these features, ranging from the tiniest convective structure
being moved by the jet streams, to the largest vortices
\citep[e.g.][]{WinJUPOSRedSpot}.  The charts reveal the dynamic interactions
within the jovian weather layer, and the long-term stability of their zonal jets
(see e.g., the regular bulletins provided by the Jupiter section of the British
Astronomical Association). The positions can be extrapolated forward in time,
enabling targeted observations by professional observatories or even visiting
spacecraft. The Juno mission, scheduled to arrive at Jupiter in 2016, is reliant
on the citizen observer community to provide this sort of contextual mapping for
the close-in observations from the orbiter.  This long-term record of Jupiter's
visible appearance by citizen scientists has proven to be an  invaluable
resource for jovian atmospheric scientists.


\CaseStudy{Time domain astronomy: Supernova Zoo and Planet Hunters}  
\label{SNZoo} 
The three defining characteristics of ``Big Data'' have come to be accepted as
volume, velocity and variety.  Time-domain astronomy projects, that indeed
require the immediate inspection of challenging volumes of live, high velocity,
complex data, can benefit from citizen science, as shown by two recent projects,
Supernova Zoo and Planet Hunters.   While transients such as supernovae or
asteroids can often be found through the use of automatic routines, visual
inspection is still used by many professional science teams as part of their
process of selecting candidates for follow-up. 

The most successful attempt to use crowd-sourcing to attack these problems to
date has been the offshoot of Galaxy Zoo described in \citet{SmithSN}.  Data
from the Palomar Transient Factory \citep{LawPTF} was automatically processed
and images of candidate supernovae uploaded on a nightly basis; this triggered
an email to volunteers who, upon responding, were shown the new image, a
reference image and the difference between the two. By analyzing the answers
given by the volunteers to a series of questions, candidates were sorted into 
three categories, roughly corresponding to ``probable supernova,'' ``likely
astrophysical but non-supernova transient'' and ``artifact.'' The results were
displayed on a webpage and used by the science team to select targets for
follow-up. Despite the Supernova Zoo site attracting many fewer classifiers 
than Galaxy Zoo, it was highly effective in sorting through data,  with
consensus typically reached on all images within 15 minutes of the initial email
being sent. 

The large dataset generated by this project was used by \citet{Brink} to develop
a supervised learning approach to automatic classification for PTF transients.
The performance of this routine, which for a false-positive rate of 1\% is more
than 90\% complete, depends on the kind of large training set that can be
generated by crowds of inspectors; this suggests a future path for large surveys
in which citizen science provides initial, training data and is followed by
machines taking on the remaining bulk of the work. Encouragingly,
\citeauthor{Brink}'s method, which makes use of a set of 42 features extracted
from survey images, has performance which is insensitive to a small fraction of
mislabeled training data, suggesting that the requirements for accuracy of
citizen science projects which aim to calibrate later machine learning may be
less stringent than otherwise thought. 

A different approach to crowd-sourced classification in time-domain astronomy is
exemplified by the Planet Hunters project, which asks volunteers to examine
light curves drawn from the dataset provided by the \emph{Kepler} mission in
order to identify interesting events in retrospect. While the task of
identifying transits from extrasolar planets is, at first glance, one which
seems more suited for automated than for human analysis, the success of Planet
Hunters in identifying more than fifty planet candidates missed by the automatic
routines suggests that there remains a role for inspection by eye in cases where
the relevant science requires samples of high completeness. Several of the
planets found by Planet Hunters are unusual: PH1b, the project's first confirmed
planet \citep{Schwamb++2013} and a circumbinary, is the first planet known in a
four-star system. Others, including the more than forty candidates identified in
the habitable zone of their parent star by \citep{Wang++2013}, might have been
expected to be recovered by more conventional searches. Planet Hunters,
therefore, is acting as an independent test of the Kepler pipeline's efficiency
\citep{Schwamb++2012} and has inspired improvements in subsequent analysis
\citep{Batalha++2013}.


\subsection{Classification Analysis}
\label{sec:class:analysis}

In most visual classification projects, working on archived image data with
little time pressure, the random assignment of task to classifier, followed by
simple, democratic treatment of the classifications has been judged sufficient.
However, the need for rapid processing of images in time domain astronomy
projects has prompted the investigation of more efficient analyses of the
classification data.  Using the Supernova Zoo project's archive as a test,
\citet{Simpson++2012IBCC} developed a Bayesian method, IBCC, for assessing
classifier performance; in this view, each classification provides information
both about the subject of the classification and about the classifier
themselves. Classifier performance given subject properties can thus be
predicted and an optimum set of task assignments calculated. 
Moreover, work by Simpson et al., as well as \citet{Kamar} and \citet{Waterhouse} on
Galaxy Zoo data, suggests that accuracy can be maintained with as few as 30\% of
classifications. 
This sort of optimization will be increasingly important for online citizen
science, especially in projects that use a live stream of data, rather than
an archive, since the classification analysis will need to be done in real time.



\CaseStudy{Rare event detection: Space Warps} 
Steps towards real-time classification analysis have been taken in the Space
Warps project.\footnote{\url{http://spacewarps.org}} 
Space Warps is a rare object search: volunteers are shown deep
sky survey images and asked to mark features that look as though they are
gravitationally lensed galaxies or quasars (Marshall et al, More et al in prep).
Extensive training is
provided via an ongoing tutorial that includes simulated lenses and known
non-lenses, and immediate pop-up feedback as to whether these training images
were correctly classified. Because real lenses are rare (appearing once every
$10^{2-4}$ images, depending on the dataset), the primary goal is to reject the
multitude of uninteresting images so that new ones can be inspected -- and this
drives the need for efficiency. Marshall et al (in preparation) derived a
simplified version of the IBCC classification analysis that updates a
probablistic model of both the subjects and the agents that represent the
classifiers in a statistically online manner. This
analysis was run daily during each of the Space Warps projects, and subjects
retired from the stream as they crossed a low probability threshold. This
algorithm is being implemented into the web application itself for future
datasets. 

The increased  efficiency of visual classification projects that will come with
real-time analysis will enable feedback on the projects' progress to be given
much more promptly -- an important part of the collaboration between
professionals and amateurs in crowd-sourcing projects.


\subsection{Visual Classification in Other Fields}
\label{sec:class:non-astro}

Although, as described in the previous section, astronomical analysis led the
development of citizen science as a data analysis tool, it has quickly been
adopted by other fields. In some cases, this adoption has been explicit. The
tools developed for Stardust@home were developed into a general purpose
library for citizen science, BOSSA.
Both this and the Zooniverse platform (which hosts many of the examples
described above) support projects from fields as diverse as ecology and
papyrology. This diversity allows general lessons about project
design to be drawn; indeed, this is an active area of research for academic
fields as diverse as computer science, economics and social science. A recent paper
by \citet{Crowston}, for example, compares Planet Hunters and Seafloor Explorer, 
a Zooniverse project which explores the health of fisheries off the coast of
North America.

Projects from other fields can also suggest strategies which could be adopted
by future citizen astronomy projects. In particular, future projects involving
analysis of survey data which has been collected for a multitude of purposes
may require a more sophisticated model for data analysis than the simple
decision tree presented by projects such as Galaxy Zoo. 


\CaseStudy{Snapshot Serengeti}  
This Zooniverse project invites the visual classification of animals in
photographs from more than two hundred motion-sensitive ``camera traps''
installed in the Serengeti National Park, and enables a particularly
sophisticated volunteer response. Driven in part by the need for an interface
which allows volunteers to state the obvious (for example, identifying
elephants, lion or zebra) and also to provide more obscure classification (for
example, distinguishing between different species of gazelle),  a variety of
classification paths are presented. In addition to just clicking buttons
identifying species, volunteers can opt for a decision tree-like approach, or
choose from a variety of similar species (``Looks like an antelope...'') or
search the descriptions provided in order to make an informed  classification
(``Show me all animals whose descriptions involve 'ears' ''). This hybrid model
has proved successful not only in encouraging classification, but also in
encouraging learning; over a Snapshot Serengeti classifier's ``career'' they are
increasingly likely to chose more direct routes.



\CaseStudy{Visual inspection of 3-D biological scans: Eyewire.}  
Another aspect of project strategy, and design, relates to the engagement of the
volunteers. The online citizen astronomy projects developed so far have tended
to emphasise co-operation between volunteers, and the results being due to a
team effort. Elsewhere, experiments with a more competitive approach to citizen
science, ``gamifying'' the activity, have been performed.

The Eyewire project\footnote{www.eyewire.org}, based at MIT, seeks to supplement machine learning
identification of neurons in three-dimensional scans. Notably, this project
incorporated some ``gamified'' elements into its design. Participants in the
project, who are asked to identify connected regions throughout a
three-dimensional scan, earn points based on participation and also have a
separate, publicly visible, accuracy score.  In addition to overall leader
boards, the project also runs short challenges including a regular Friday
``happy hour'' in which participants compete on specific problems. Eyewire is
also notable for its other strong community elements, with a chat room open and
available to all participants in the project (supplemented, incidentally, by a
``bot'' built by a participant which answers frequently asked questions from new
users). Its first result, which drew on mapping of so-called `starburst' neurons, was 
published in mid-2014 \citep{KimEyewire}.


\section{DATA MODELLING}
\label{sec:model}

New understanding of the world comes from the interpretation -- fitting -- of
data with a physical model. Such ``data modelling'' often involves technical
difficulties that computers may find hard to overcome, associated with complex
and/or computationally expensive likelihood functions. Humans, by applying their
developed intuition, can contribute a great deal to the exploration of a model's
parameter space by closing in quickly on those configurations that fit the data
well. This process can be particularly satisfying, rather like solving a puzzle.
Meanwhile, many ``machine learning'' techniques effective in one field can often
be adapted to astronomical problems: there are plenty of citizens with the
skills to do this. How have citizen scientists been involved in model making and
data fitting in astronomy, and other fields, to date?


\CaseStudy{The Milky Way Project}
\citet{Simpson++2012MWP} provided volunteers with a fairly flexible set of
annulus-drawing tools, for annotating circularly-symmetric ``bubble'' features
in colour-composite (24.0, 8.0 and  4.5$\mu$m) infrared images from surveys
carried out by the Spitzer space telescope. These bubbles are hypothesized  to
have been caused by recently-formed high mass stars at the centre each. The
(bubble) model in this case is simple and recognizable, making both the
interface construction and its operation relatively straightforward. The large
sample of  bubble models have been used to investigate the possibility of
further star formation being triggered at the bubble surfaces
\citep{KendrewEtal2012}. A subsequent effort \citep{Beaumont} used data provided
by the project to train a machine learning algorithm, \emph{Brut}, in bubble
finding. \emph{Brut} is able to identify a small number of sources  which were
not identified in the \citeauthor{Simpson++2012MWP} catalog. These bubbles were
difficult for humans to identify,  owing to their lying close to bright sources,
and so having low contrast relative to their surroundings.

In addition, \emph{Brut} has proved effective at identifying suspect bubbles
included in the previous (pre-citizen) surveys. Given the relatively small size
of the MWP sample, the main use of machine learning here has been to provide an
independent check on the citizen classification data; for larger samples, as
discussed below, an approach in which machine learning is trained on citizen
science data, and gradually takes over the classification task could be
considered. 


\CaseStudy{Modelling Lens Candidates} 
The Space Warps project (\Sref{sec:class:analysis}) has an informal data
modeling element. The classification interface is restricted to enabling
identification of candidate gravitationally-lensed features, but all the images
are available via the project's discussion forum. A small team of volunteers
(including several citizens who helped design the project) has engaged in
modeling some of the identified lens candidates using web-based software
developed and supported by the project science
team.\footnote{\url{http://mite.physik.uzh.ch}} Results from a small test
program show that the model parameters derived by the  ensemble of citizens are
as accurate as those derived by experts 
(Kueng et al, in prep).
A pilot collaborative modeling analysis was carried out and written up by a
small group of Space Warps volunteers\footnote{See
\url{http://talk.spacewarps.org/\#/boards/BSW0000006/discussions/DSW00008fr} for
the forum thread that was used.} \citep{Wilcox2014}.



\CaseStudy{Galaxy Zoo: Mergers} 
This has been perhaps the most advanced attempt at data modeling in 
astronomical web-based citizen science \citep{HolincheckEtal2010,WallinEtal2010}.
Here, simple N-body simulations of galaxy mergers were performed in a java
applet, and the results selected according to visual similarity to images of
galaxy mergers (previously identified in the Galaxy Zoo project). A key
hypothesis here is that the inspectors of the simulation outputs would be able
to find matches to the data more readily than a computer could, for two reasons.
First is that humans are good at {\it vague} pattern matching: they do not get
distracted by detailed pixel value comparisons but instead have an intuitive
understanding of when one object is ``like'' another. The second is that
initializing a galaxy merger simulation requires a large number of parameters to
be set -- and its this high dimensionality  that makes the space of possible
models hard to explore for a machine. Humans should be able to navigate the
space using their intuition, which is partly physical and partly learned from
experience gained from playing with the system. Initial tests on the marging
system Arp 86 showed
the crowd converging on a single location in parameter space, and that the
simulated mergers at this location do indeed strongly resemble the Arp 86
system. The authors have since collected thousands of citizen-generated models
for a sample of a large number SDSS merging systems (Holincheck et al, in
preparation). 

\begin{figure}[!ht]
\centering\includegraphics[width=\linewidth]{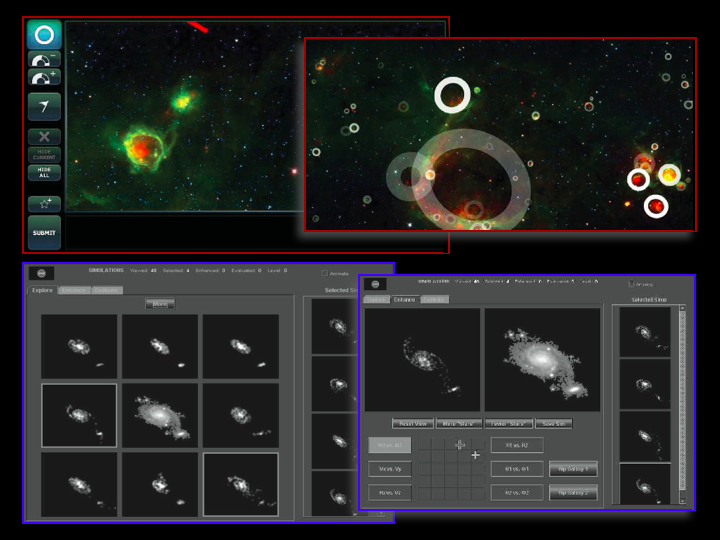}
\caption{Examples of image modeling in web-based citizen science projects. Top
row: star formation ``bubble'' identification and interpretation in Spitzer
images in the Milky Way Project, with the annotation interface shown on the
left, and some example (selected, averaged) bubbles on the right. Images from
\citet{Simpson++2012MWP}. Bottom row: matching N-body simulated merging
galaxies to SDSS images in the Galaxy Zoo Mergers project (left), and
exploring parameter space two parameters at a time to refine the models
(right). Screenshots from \citet{HolincheckEtal2010}.}
\label{fig:modeling}
\end{figure}


\CaseStudy{Protein Modeling with Foldit}  
One of the most successful examples of crowd-sourced, ``manual'' data modeling is the
online multi-player 3-D protein modeling game,  Foldit
\citep{Cooper++2010}\footnote{\url{http://fold.it}} In this pioneering
project, players compete in teams to find the best -- lowest free energy --
molecular structures for particular protein ``puzzles.'' 
These puzzles are naturally visualizable in three dimensions, but 
they nevertheless involve thousands of degrees of freedom, in a parameter
space that is notoriously hard to explore.
Under the hood is the
professional Rosetta structure prediction methodology; the player's scores are
simply the negative of the Rosetta-computed energy. Foldit provides an
accessible interface to the Rosetta toolkit, which provides multiple ways to
interact with the protein structure as the global minimum energy solution is
sought. The Rosetta model parameter free energy hyper-surface is completely analogous
to the complex likelihood surface of any non-linear model, the kind of model
that is to be found in planetary system dynamics, gravitational lenses,
merging galaxies, and many other astrophysical data analysis situations.

Results from Foldit have been very encouraging, with the players discovering
several new protein configurations, leading to improved enzyme performance 
\citep{Eiben++2012} and new understanding of retroviral drug design
\citep{Khatib++2011a}. The team have suggested several features of Foldit that
appear to them to have underpinned its success. Recipes for manipulating the
protein structures (that codify strategies) can be shared within teams, and
later made available by the Foldit team to the whole community -- these
algorithms evolve rapidly as different players modify them, and can rival (if
not out-perform) strategies developed by professional scientists 
\citep{Khatib++2011b}. The game provides multiple sources of motivation
(competition between players, collaboration within a team, short term scores,
long term status) which appeal to a variety of players.


\CaseStudy{Online Data Challenges}
We now turn to data modeling by citizens implementing machine learning
techniques in astronomy, via analysis challenges organised by members of the
professional astrophysics community. The measurement of weak gravitational
lensing by large scale structure (``cosmic shear'') relies on the measurement of
the shapes of distant, faint galaxies with extreme accuracy. The STEP
\citep{HeymansEtal2006,MasseyEtal2007} and GREAT
\citep{BridleEtal2010,KitchingEtal2012,KitchingEtal2013a} blind galaxy shape
estimation challenges have had an enormous impact on the field, revealing biases
present in existing techniques, and providing a way for researchers outside the
world of professional cosmology to participate. In particular, the GREAT08
challenge saw very successful entries from two (out of a total of 11) teams of
researchers from outside of astronomy (albeit still professional researchers),
including the winner. A companion, somewhat streamlined galaxy shape measurement
challenge, ``Mapping Dark Matter,'' which was hosted at the Kaggle
website\footnote{\url{http://www.kaggle.com/c/mdm}} \citep{KitchingEtal2013b}.
The wider reach of this platform led to over 70 teams making over 700 entries to
the competition; many of the teams did not contain professional astronomers,
although most were still from academia.

In a comparison with the GREAT challenges, the \citeauthor{KitchingEtal2013b}
found a factor of several improvement in shear accuracy over comparable previous
challenges, and suggested two interesting explanations for this success. First,
the challenge was designed to be as accessible as possible, with an extensive
training set of data that needed very little explanation; in this way the
challenge was geared towards {\it idea generation}. Second, they noted that the
competitive  nature of the challenge (a webpage leaderboard was updated in real
time as entries were submitted) seemed to stimulate the analysts into improving
their submissions. Kaggle offers cash prizes, which will have had some effect as
well (the pot was \$3000 for this challenge, even if indirectly).

Two more astronomical Kaggle challenges have since been set. The first
involved inferring the positions of dark matter halos based on their weak
lensing effects
\citep{Harvey++2014}\footnote{\url{http://www.kaggle.com/c/DarkWorlds}}  This
challenge attracted the attention of 357 teams, perhaps due to its larger
prizes, and led to an improvement in halo position accuracy of 30\%.  It also
sparked some debate in its forums as to the design of the challenge: the
models used to generate the data, the size of the test datasets (and
consequent stability of the leaderboard),  the choice of leaderboard metric
and so on. These issues are also of generic importance for scientists looking
to crowd-source algorithm development. It is interesting to note that the
Kaggle forums are a useful resource for the Kaggle development team: the
citizens who are active there do influence the design of the site
infrastructure and challenge rules (D.~Harvey, priv.~comm.). The most recent
Kaggle astronomy challenge was to reproduce the Galaxy Zoo crowd-sourced
galaxy morphologies, given the same color
images.\footnote{\url{http://www.kaggle.com/c/galaxy-zoo-the-galaxy-challenge}}



Like Foldit's ``recipes,'' the Kaggle challenges are crowd-sourcing the
development of new algorithms. As data science plays an increasingly important
role in industry and commerce, we can expect the number of citizens interested
in applying their skills to science problems in their spare time to grow. The
challenge is to present those problems in meaningful ways, to enable high value
contributions to be made. While members of this community may not identify as
``citizen astronomers,'' there is clearly an opportunity for citizen data
scientists to play an important support role.


\section{CITIZEN-LED ENQUIRY}
\label{sec:explore}

The previous sections have focused on specific, and somewhat  isolated
activities in which citizens have participated. In most cases, the community's
involvement has been a {\it contribution} to a scientific investigation defined
by professionals. The most important part of any scientific investigation is the
question at its heart: what is it we are trying to find out about the universe?
In this section we look at some cases where the process of enquiry, the science
itself, has been instigated or led by citizens.  

In principle, this is an area of great potential. The constraints of funding
proposals and management of research groups can often mean that professional
scientists focus very narrowly on particular topics of research, specializing in
particular techniques or datasets.  Steering away from this course implies
taking risks with time management, and allocation of resources to an ultimately
fruitless research area can be detrimental to careers.  Citizen scientists are
largely free of these managerial and budgetary constraints, and are able to
devote their attentions to whatever topics interest them. Moreover, we might
expect outsiders to ask some unusual questions, and make connections and
suggestions that highly focused professionals may not have thought of. What are
some enquiries that citizens have led in astronomy to date, and how have they
been enabled and supported?


\CaseStudy{Saturn Storm Watch.}  
In this project, Cassini's observations of lightning emissions were connected to
active amateur observations of the convective cloud structures within the giant
planet atmosphere \citep{11fischer},  and the vertices of Saturn's bizarre
north polar hexagon \citep{godfrey88} (a 6-sided planet-encircling wave that has
persisted for at least 30 years but that has only recently been observed, by
amateur astronomers).  In the first case, citizen scientists wished to identify
the source of the radio emission detected by Cassini, after being alerted to
them on the Planetary Virtual  Observatory and Laboratory.
\footnote{\url{http://www.pvol.ehu.es}} In the latter case, the long-term
evolution of the hexagon vertices was used to understand what sort of wave this
is, and to identify its origins.



\CaseStudy{The Galaxy Zoo Forum.} 

The best known serendipitous discovery emerging from the Galaxy Zoo project is
``Hanny's Voorwerp'' \citep{Lintott++2009}, a galaxy-scale light echo which
reveals a recent ($\sim 100,000$ years ago)  shutdown of AGN activity in IC
2497, a neighboring spiral galaxy \citep{Keel++2012}. 
The discovery of the Voorwerp was first
recorded in the Galaxy Zoo forum a few weeks after the project started, and
inspired a more systematic search for similar phenomena in other galaxies. This
project, made possible by the deep engagement in the forum community of Galaxy
Zoo science team member Bill Keel, succeeded in finding more than forty
instances of clouds which appear to have been ionized by AGN activity, in
systems a third of which show signs of significant drops in activity on a
timescale of tens of thousands of years.  

The ability of the Zoo volunteers to carry out their own research, moving far
beyond the mere ``clockwork'' required by the main interface, is best
illustrated by the discovery of the Galaxy Zoo Green Peas
\citep{Cardamone++2009}. These small, round and, in SDSS imaging, green systems
are dwarf galaxies with specific star formation rates (SFR per unit mass) which
are unprecedented in the local Universe, matched only by high-redshift
Lyman-break galaxies. Volunteers not only identified these systems, but
organized a systematic search and further review of them. This effort included
the use of tools designed by SDSS for professional astronomers to acquire and
study spectroscopic data. 

While the discovery of the Peas demonstrates the exploration ability of the
Galaxy Zoo citizen community, it is important to note that the simpler, initial
interaction provided by the main classification interface was necessary in order
to develop that community in the first place. The participants in the citizen
scientists' investigation of the Peas did not arrive on the site wanting to dig
into spectra or confident of their ability to do so; these were the results of
their participation. The project acted as an ``engine of motivation'' in
inspiring its participants to become more involved. 


\CaseStudy{Lightcurve analysis on Planet Hunters {\it talk}.}  

The data modelling examples of \Sref{sec:model} all involved modeling
infrastructure provided by either the project's developers or their science
teams. Planet Hunters provides a case where citizens have carried out their own
modeling analysis, using their own tools. Critical to this endeavour was the
ability of a small,  and increasingly expert, group of volunteers to identify
objects worthy of further analysis. For Galaxy Zoo, the forum had served this
purpose but, as as the project matured, participation in discussions became
restricted to a small and decreasing fraction of the community. Planet Hunters
was the first Zooniverse project to introduce an integrated discussion tool,
known as \emph{talk}. Classifiers were asked, after viewing each light curve,
whether they wanted to discuss what they had seen; more advanced users could
then harvest interesting candidates from these posts. For example, the
candidates presented in \citet{LintottPH} were initially collated by volunteers.

Their involvement was not limited to collecting Planet Hunters candidates.
Making use of the Kepler archive, these advanced users were able to investigate
the full set of data for candidate stars, producing periodograms and making
fits  to transits to derive planet candidate properties. Some of this analysis,
for example checking the Kepler field for  background sources, can be carried
out online with tools originally intended for professional astronomers, but much
was done off line using Excel or other software.\footnote{The expense of IDL
licenses was a major barrier to further modelling; much  software used by the
Kepler team is written in this proprietary language.} Nor was this sort of work
restricted to planet candidates; interesting variable stars, including several
new RR Lyrae systems, and cataclysmic variables \citep[e.g.\ ][]{KatoOsaki} 
have been discovered and analysed by Planet Hunters volunteers.


\CaseStudy{Galaxy Zoo: Quench.} 
Examples such as those above show that advanced work is possible within
distributed citizen science projects, but that this requires volunteers to take
on such tasks themselves. In order to increase the number, and perhaps the
diversity, of volunteers moving beyond simple classification, experiments have
been conducted to provide more scaffolded experiences. One of the most ambitious
was the Galaxy Zoo: Quench project (Trouille et al. in prep) which offered
volunteers the opportunity to ``experience science from beginning to end.''

In this project, classification of a sample of potential post-merger galaxies
selected from the main Galaxy Zoo sample was followed by open exploration of
both the classification data and the metadata for these galaxies (available from
the Sloan Digital Sky Survey) by the volunteers, enabled by a
``dashboard\footnote{\url{http://tools.zooniverse.org/\#/dashboards/galaxy\_zoo}}.''
Thousands of users (around 20\% of those who participated in the classification
stage and discussion) led to the formulation of a set of astrophysical
interesting conclusions; a small number of participants (~10) collaborated on
writing a paper (in preparation). These later stages required intensive support
from professional scientists (and in fact it was constraints on their time that
prevented earlier submission of the paper). 

Quench demonstrated that a hierarchical approach, with simple tasks leading to
more advanced analysis, can be successful in encouraging large numbers of
volunteers to move beyond simple classification; the number participating in
exploring the data was much higher as a percentage of participants than in
Planet Hunters. However, once engagement with the literature (either by reading
or writing) is required there remains no substitute for significant involvement
by professionals.

%


\section{UNDERSTANDING THE CITIZENS}
\label{sec:crowd}

Having surveyed some of the activities involving citizen scientists, we can
now consider some questions about this community itself. Who participates in
citizen science, and what motivates them?


\subsection{Demographics}
\label{sec:crowd:demographics}

Who is participating in citizen astronomy? We might expect the demographics to
vary with activity, and with the level of commitment required. We have some
understanding of at least the former division from two studies that were
carried out approximately simultaneously, one of the community  participating
in Galaxy Zoo, and another of the American Association of Variable Star
Observers (AAVSO).  \citet{Rad++2013} surveyed the Galaxy Zoo volunteer
community to investigate their motivations (\Sref{sec:crowd:motivation}
below), via a voluntary online questionnaire. The 11,000 self-selected Galaxy
Zoo users identified as 80\% male, with both genders having an approximately
uniform distribution in age between their mid-twenties and late fifties. The
authors point out that this is close to the US internet user age distribution,
except for slight but significant excesses in numbers of post-50s males,
post-retirement people of both genders, and a deficit in males under 30. The
survey respondents  also tended to be more highly educated than average US
internet users, with most holding at least an undergraduate degree, and around
a quarter having a masters or doctorate. Very similar findings were reported
by \citet{COSMOQUESTsurvey} from a survey of COSMOQUEST project participants.

These findings can be compared with a survey of the members of AAVSO:
\citet{P+P2012} received over 600 responses (corresponding to about a quarter
of the society's members). The education levels of
the AAVSO repondents matches the Galaxy Zoo community very closely; the AAVSO
age distribution is more peaked (in the mid fifties), with a similar post-60
decline but also a marked absence of younger people. The online nature of the
Galaxy Zoo project seems to have increased the participation of younger (pre
middle-age) people. Likewise, the Galaxy Zoo gender bias, while itself
extreme, is less so than at AAVSO, where some 92\% of survey respondents were
male. One additional piece of information provided by the AAVSO survey is the
profession of the variable star observers: most (nearly 60\%) of the survey
respondents were found to be working in science, computer science, engineering
and education. 

The Galaxy Zoo and AAVSO communities differ by more than just the nature of
their activity. The smaller AAVSO community is arguably more engaged in its
research, in the sense that a larger fraction of its membership is active in
taking observations and contributing to analyses. It would be very interesting
to know how citizen scientist motivation varied with the level of
participation: dividing the Galaxy Zoo community into volunteers that
contribute to the  forum and those who don't could be interesting; perhaps
more so would be to repeat the analysis of \citeauthor{Rad++2013} over a wide
range of projects, and look for trends there. The emergent picture thus far,
however, is of a well-educated (and often scientifically trained)  but
male-dominated citizen science community, whose female and younger membership
is likely to have been, at least in part, enabled via projects being hosted
online. Continuing to lower the barriers to entry for currently
under-represented demographic groups would seem both important, and within
reach.


\subsection{Motivation}
\label{sec:crowd:motivation}


What motivates citizen scientists? The two demographic studies referred to above
also covered this question; having previously \citep{Rad++2010} identified 12
categories of motivation in an earlier pilot study, \citet{Rad++2013} asked the
170,000 Galaxy Zoon volunteers at the time to comment on how motivated they were
by each of these categories, and which was their primary motivation. The 6\% who
responded gave consistent answers to those given by around 900 forum users who
responded in a separate appeal, allowing conclusions about this presumably more
engaged sub-population to be drawn. A desire to {\it contribute} to science was
found to be the dominant primary motivation, being selected by 40\% of
respondents. {\it Astronomy}, {\it science}, {\it vastness}, {\it beauty} and 
{\it discovery} were all motivation categories that were found to very important
to the volunteers, while {\it fun}, {\it learning} and {\it community} were less
important. 

The AAVSO demographic survey \citep{P+P2012} found similar results: over a third
of variable star observers cited {\it involvement in science and research} as
their primary source of motivation. However, a similar number gave an {\it
interest in variable stars} as theirs, perhaps reflecting a stronger focus on
the science questions involved than is present in the Galaxy Zoo community. Both
groups of citizen scientists are clearly quite serious in their reasons for
taking part: their motivations are actually very close to those of professional
scientists, as many readers of this review will recognize. Perhaps surprisingly,
the participants in online data analysis citizen science projects seem to a
large extent to be a distinct community from those who participate in more
traditional amateur astronomical activities. Galaxy Zoo classifiers, for
example, are not, for the most part, regular amateur observers. 

While research on the skill, and conceptual understanding, that  people aquire
while participating in citzien science activities is still in its early stages,
there are some hints that continued engagement is correlated with both
performance in the task at hand, and understanding of the physics and astronomy
underlying the task. \citet{Prather++2013} offered Galaxy Zoo and Moon Zoo
volunteers the opportunity to take questionnaires that tested their
understanding of the astrophysics associated with each project, and found that
performance on this questionnaire correlated with high levels of participation
in the projects. In the Space Warps project, the probabilistic model for the
crowd includes a measure of each classifier's skill; a strong correlation is
seen between a classifier's skill, and the number of images they have seen
(Marshall et al, in prep.). It seems as though the skillful classifiers remain
engaged in the project for a long time, while almost no long-term participants
have low skill -- an observation consistent with the volunteers being motivated
by contributing to science.


\subsection{Competition or Collaboration?}
\label{sec:crowd:gamification}

As seen in \Sref{sec:class:non-astro} and \Sref{sec:model}  above,
non-astronomical projects may have much to teach us about ``gamification'' as a
motivator -- the inclusion, either explicitly or implicitly, of game-like
mechanics such as scores, ``badges'' or other rewards, leader boards, and so on.
The Foldit team present a strong case for games as drivers of activity in
citizen science, and the Kaggle challenges depend on competition to stimulate
engagement. What has been the experience in citizen astronomy so far?

An early experiment with Galaxy Zoo showed that the addition of a score
de-incentivised poor classifiers, but also resulted in the best classifiers
leaving, presumably having been satisfied once a top score was achieved. A
recent study by \citep{Eveleigh++2013} of the Zooniverse's Old Weather
project, which included basic rankings for classifiers, highlighted these
dangers, identifying volunteers who were alienated by the addition of this
game-like score.They felt discouraged when top scores could not be matched, and
worried about data quality if the scoring scheme rewarded quantity of
classifications rather than accuracy. Taking seriously the above finding  that
citizen scientists are motivated by a perception of authentic participation in
research, it seems right to be cautious about introducing elements which are, or
which are perceived to be, in tension with this primary motivation. 

Furthermore, the introduction of a significant incentivizing scheme relies on an
accurate model of what ``correct'' behaviour would look like. This may prove to
be a significant barrier to accuracy if such a model is not available. For
example, in Planet Hunters, such a model would not have included unusual systems
such as PH1b. Where a strong incentive scheme results in near-uniform classifier
behaviour, a loss of flexibility in later data analysis could be incurred.  A
strong comparison of the type of reward structure utilized by Eyewire and the
approach used by projects such as Galaxy Zoo is needed, in order to inform
future project design. 

The surveys described in the previous section reveal a community of people many
of whom may have left academic science behind as soon as they finished their 
education, but who still maintain a passion for astronomy and the boundaries of
knowledge.  Their thirst for new information, and the  desire to be part of the 
scientific process drives them to actively observe the  night sky or to
participate in analysis of large datasets.  For the more motivated people
involved in citizen science, being part of a community,  albeit a distributed
one, brings great enjoyment and satisfaction.  With the connectivity brought by
the internet, there is a social  aspect of citizen science that unites people
with this shared interest, which may be far removed from their ``normal'' 
lives. However, {\it community} was not found to be a strong motivator for the
Galaxy Zoo volunteers -- but it is nevertheless very important for the Galaxy
Zoo forum users. More recent Zooniverse projects have sought to widen
participation in community discussion, hypothesizing not that it will more
strongly motivate people, but because it will help them make better
contributions. Tests of hypotheses like this should be helpful in guiding
citizen science project design.


\section{THE FUTURE OF CITIZEN ASTRONOMY}
\label{sec:future}


During this review a picture has emerged of two types of very active and engaged
citizen astronomy community, which we might label observers and classifiers. 
Although theses communities come together in differing ways (by self-assembly
through local groups linked by national and international networks, or by
joining online projects built by professional organisations), they have reached
a similar degree of internet-enabled connectedness, both with each other and
with the groups of professional astronomers with whom they  collaborate. They
also share the common motivation of being involved in, and contributing to,
science. In this section we look ahead, to the next decade or so, and discuss
the likely paths that citizen astronomy will take, as the available technology
advances and professional astronomy evolves. In it we try to identify the niches
that citizens might best occupy in this changing environment, and also some key
challenges that those who find themselves planning citizen science projects are
likely to have to face.



\subsection{The Future of Citizen Observing}

In professional astronomy, the wide field survey era is upon us: SDSS provided
the data for Galaxy Zoo, and other, larger surveys are planned or underway. Key
science drivers for projects such as LSST and the Square Kilometer Array include
mapping cosmological structure back into the reionisation era, and  further
opening the time domain; these will yield datasets of significantly increased
volume, throughput rates, and complexity.  Follow up observations of new
discoveries made at greater depths will be made with giant facilities such as
ALMA and the various planned Extremely Large Telescopes, while distributed
arrays of robotic telescopes, operating in remote regions with excellent
atmospheric conditions, and trained to observe a target in a regular fashion
over multiple nights will be able to take advantage of wealth of new transient
phenomena. 

These future advances in technology may in one sense widen the gap between
citizen scientists and professionals again. For example, networked telescopes
capable of quasi-continuous observations over 24 hour periods could be used to
develop a consistent high-quality dataset for cloud tracking on Venus, Mars or
the giant planets; as the images would be homogenous, we can envisage automated
software identifying morphological peculiarities over time, replacing the
crowd-sourced citizen analysis currently underway.  However, such an investment
would require both international funding and considerable time and effort: the
availability of citizen observers will remain a factor.



However, the advances in hardware becoming available to citizen observers
suggest other roles they could play. Larger optics, more sensitive cameras, and
spectral coverage extending to longer wavelengths in the infrared could permit
citizen investigations of Uranus and Neptune, the Trans-Neptunian and Kuiper
Belt objects, and a wider variety of bright variable stars.  Transits of
extrasolar planets in front of their parent stars would be permitted from modest
observatories provided they had stable conditions.  New platforms might also
become available to the citizen scientist, including balloon-borne observatories
that provide crisper and more detailed observations of astronomical targets. We
may well see networks of citizen deep sky observers investigating new bright
transients found in the wide field surveys,  while continuing to expand their
own surveys.

Aside from pushing the observational boundaries, one challenge that amateur
astronomy may face is its own big data problem.   For example, solar system
video monitoring projects are likely to need automated feature detection of some
kind; other observing campaigns may also generate more data than is easily
manipulated. Will this community take to crowd-sourcing its visual inspection?
The Zooniverse platform is currently being redeveloped to enable easy upload of images
and launch of projects; such a facility may be used by citizen scientists as well as by 
professionals. 



%

\subsection{The Future of Crowd-sourced Visual Classification}

The point at which human review of data is no longer necessary has been forecast
for decades, but as we have seen above, the number of problems for which manual
review of images or data is still carried out is considerable. Even if the
proportion of data for which human inspection is necessary decreases
dramatically over the next decade (due to advances in automatic analyses), the
continued growth in the size of astronomical datasets should ensure that there
remains plenty for citizen scientists to do. 

Consider the example of optical transients. The LSST system overview paper
\citep{LSSTsystem} gives a conservative estimate of $10^5-10^6$ alerts per
night. Even if, after automated brokerage, only 1\% of these require human
classification, then that still might lead to 10,000 objects requiring
inspection and interpretation every night -- roughly one every 100 seconds.
Given the increased reliability, and likelihood of serendipitous discovery,
provided by citizen inspection, we should take seriously the incorporation of
open inspection into plans for LSST transients. Similar arguments (with large
error bars) can be made for other surveys: inspection of transients for LOFAR
already requires some human intervention \citep{LOFAR}.

Implicit in this way of thinking is the sharing of work between human and
machine classifiers.  A simple example of human-machine task allocation was
mentioned in \Sref{sec:class:analysis}, where machine analysis of PTF images
identified those that contained candidate supernovae needing inspection by
volunteers. It is worth noting that the inclusion of human inspection changed
the nature of the machine learning task; instead of optimising for purity
(producing a small but accurately classified set of candidates), the task for
machine learning became one of identifying a subset of the images which
contained many false positives but also a complete set of all supernovae. 
In this example, human and machine classification proceeded in series rather
than in parallel, but more complex interactions can be imagined. 

The accuracy of machine learning typically depends on the quality of the
training or ``gold standard'' data which can be provided for the problem in
question. Citizen science projects can assist by providing training sets which
are orders of magnitude larger than might otherwise have been available, while
work by \citet{Banerji++2010} established that the confidence intervals provided
by classifications from multiple volunteers can also improve machine learning
accuracy.  Predicting human responses (in the form of probabilities of
classification) is an easier task than straightforward sorting. We expect,
therefore, intermediate-size surveys to benefit in the future from a ``citizen
science phase,'' in which data is classified by volunteers prior to the
automation of the task. This pattern has already been followed by the PTF
supernova project discussed above, but perhaps it is more useful to think of the
citizen scientists as providing training sets on demand, so that as conditions
change from night to night, or the performance of the instrument evolves over
time, a small percentage of the total data is always processed by humans in
order to provide a constantly updated training set. 

If we are using classifications of gold standard data to assess the performance
of human classifiers, it is straightforward to include machine classification in
the same system. In this way, the task of classification could be shared
dynamically and in real time between machine and human classifiers, improving
the efficiency of the system. Significant work has already been carried out for
the nearly analogous problem of assigning tasks to an ensemble of imperfect
machine classifiers whose characteristics are known 
and for
Mechanical Turk-like systems where a fixed payment is provided for a task but
the problem of adding in volunteers is significantly harder.
For the machine-only case, each classification task can be treated as having a
known cost (perhaps the processing time necessary for a given routine), but when
assigning tasks to  volunteers, who are able to leave whenever they like, other
costs must be taken into account. In order to create a viable system, it is, in
fact, necessary to measure how {\it interesting} 
a task or set of tasks is, and this
requirement may conflict with the need for efficiency. As an example, consider a
Galaxy Zoo-like system which assigned the hardest galaxies to the best
classifiers. This would result in a steady diet of faint fuzzy objects for the
best classifiers; if they are motivated in part by the variety of images seen,
then such a system would tend to systematically drive away its best classifiers.
A study of Snapshot Serengeti even revealed that seeing a more impressive image
early in a classifier's career (as measured by the number of volunteers who
added it to their list of favourites) tended to decrease the number of
classifications received from that volunteer in the long run. Considering
individual classifications in isolation is clearly not sufficient; the entirety
of a volunteer's career must be considered when assigning tasks. We should be
wary of over-specialisation even when efficiency is paramount. Complexities like
these indicate a clear need for research into novel systems for task assignment,
in order to scale citizen science to the challenges of the next generation of
surveys.




\subsection{Advanced Citizen Activities in the Future}

As we have seen in previous sections, volunteers can and do move beyond simple
classification problems, and such behaviour could become increasingly important
as the volume and complexity of astronomical data continues to increase.  We can
imagine providing user-friendly, web-based tools enabling  fairly sophisticated
data analysis to be performed by anyone with a browser. The experience
documented above invites us to consider the possibility of teams of citizens 
performing analyses that currently require a significant amount of research
student time.  Checking survey images and catalogs  for processing failures and
fitting non-linear models to data are just two possibilities. Just as research
students adapt and develop the tools they are first presented with, the Kaggle
and Foldit experiences point strongly towards a model where citizens are also
enabled to evolve their tools. Open source tool code is a minimal requirement in
this model; finding ways beyond this to support citizen algorithm development
seems to be likely to pay off.

In terms of supporting citizen-led enquiry, an example of best practice exists
in the way that the Sloan Digital Sky Survey's \emph{sky server} provided tools
for both professional (or advanced) researchers alongside simplified versions
aimed primarily at educational use. This structure has the twin benefits of
providing near-seamless transitions from simple to more advanced interfaces, and
of providing extra pressure to make the resulting interfaces easily usable 
(something which benefits all users, not just citizen scientists!). Designers of
science user interfaces for upcoming large projects would do well to bear these
twin audiences in mind. Indeed, the more citizen-accessible the interfaces to
the upcoming public wide-field survey databases can be made, the better chance
we will give ourselves of enabling and supporting ``bottom-up'' citizen
science.  

This term, introduced by Muki Haklay and collaborators, represents an ambition to produce
citizen science projects that are driven by the participants. Moving beyond the 'top down'
structure of most astronomical citizen science projects is, as we have shown, a significant
challenge - but one that is, perhaps, worth taking on.



\subsection{The Future of Citizen Science}

As well as enabling access to the data, citizen science projects looking to
engage larger crowds of volunteers will likely face some other  challenges. We
might expect contributing to science via large international public datasets to
appeal to citizens of many nations: while translation of project materials is simple, 
coordinating a scientific \emph{discussion} across multiple language barriers could prove difficult. Having a critical mass of
professional scientists interacting in each language would seem the most
important factor. Even within a single language group, collaboration is
difficult to achieve with  very large numbers. The hierarchical system of
citizen discussion  moderators bridging the gap between science teams and the
crowd has worked well in the Zooniverse projects, although it requires
significant commitment and effort from the volunteer moderators. Access to
professional scientists can be somewhat improved by regular webcasts (as
provided, for example, by the Galaxy Zoo science team); this provides at least
partially the scientific dialogue that is most valuable to the citizen
collaborators. Certainly these can supply much-needed feedback as to the utility
of the citizens' efforts, as the professionals report on how the
citizen-provided data is being used. We might imagine regular broadcasts from
the projects providing the data as playing a significant role in motivating and
sustaining a crowd of volunteers. However, for the foreseeable future it remains the case
that astronomical surveys and other organisations will seek to use citizen science as a 
way of expanding the amount of science that can be done; systems which rely on significant
intervention from paid professionals will likely fail (or at least be a luxury available to few). A
focus on systems which can maximise scientific return and volunteer participation without
substantial intervention remains necessary. 


\section{CONCLUDING REMARKS}
\label{sec:conclusions}

Over the last two decades, citizen astronomy has undergone a period of rapid
growth, primarily due to the sharp increase in the ease with which people can
form communities and work together via the world-wide web.   A number of very
productive ``Pro-Am collaborations'' have formed, to observe a variety of bright
astronomical objects in ways that capitalise on the flexibility, availability
and skill of the amateur observing community. Professional-led visual
classification projects have appeared, attracting three orders of magnitude more
citizens to the field than were previously engaged in amateur observational
research. Citizen-classified training sets have been used to improve the
performance of  machine learning approaches, suggesting that we should think in
terms of ``human-machine partnerships.'' Citizens have been challenged to take
part in data analysis tasks of increasing sophistication and difficulty, and
experiments in professionally-guided ``bottom up'' citizen research have begun.

In this review, we have consistently seen that the best citizen science in
astronomy has come from organised communities asked to play to their strengths,
and which operate in niches insufficiently occupied by either professional observers or
automated classification software. The citizen astronomers are passionate about
the subject, and are encouragingly motivated by being of service to science. We must recognize
that a critical feature of ``citizen science'' is the enabling of amateurs to
make authentic contributions to the research topic in question: this in turn
should drive us to seek out those  tasks that cannot be done by other means. 

The observational and classification citizen scientist communities are similar
in their diversity regarding both their motivation and their ability to
contribute;  this diversity means that good citizen science projects are ones
that provide both a low barrier to entry, but that also provide (or support the
development of) tools that enable their emergent experts to maximize their
contributions to science.  Indeed, the most dedicated volunteers have proved
capable of developing and using fairly advanced astronomical techniques,
suggesting that we are likely to continue to see increasing numbers of citizens
co-authoring papers in high impact research journals. While not everyone who
takes part in a project wants to move to more advanced work, providing the
opportunity to do so is important.

Each of the case studies presented in this review has been an experiment in
citizen science: amateur and professional astronomers alike have had good ideas
for ways to make use of the public's skills and abilities, tried them out, and 
made progress in astronomy -- and in doing so revealed something about how
citizen science can work. Human potential is vast: citizen astronomy seems to us
to be an experiment well worth continuing. 


\section*{Acknowledgments}

We are most grateful to the following people for their suggestions, 
comments, and clarifications: Grischa Hahn, ... 

We thank David~W.~Hogg, Arfon Smith, Laura Whyte, ... for many useful
discussions about the practice of citizen science in astronomy.

PJM and LNF were supported by Royal Society research fellowships at the
University of Oxford. The work of PJM was also supported in part  by the U.S.
Department of Energy under contract number DE-AC02-76SF00515.
CJL acknowledges...


\section{LITERATURE CITED}

\bibliographystyle{Astronomy}




\end{document}